\pdfoutput=1
\documentclass[11pt]{article}

\usepackage[final]{acl}

\usepackage{times}
\usepackage{latexsym}

\usepackage[T1]{fontenc}
\usepackage[utf8]{inputenc}

\usepackage{microtype}

\usepackage{inconsolata}

\usepackage{graphicx}
\usepackage[most]{tcolorbox}
\usepackage{multirow}
\usepackage{graphicx} % 插入图片
\usepackage{subcaption} % 支持子图
\usepackage{adjustbox}
\usepackage{xcolor}
\usepackage{amssymb}
\usepackage{amsfonts}
\usepackage{xspace}
\usepackage{amsmath}
\usepackage{booktabs}
\usepackage{graphicx} 

\tcbuselibrary{skins, breakable} % 加载必要的 tcolorbox 库
\usepackage{enumitem} % 用于控制 itemize 的间距

\newcommand{\model}{CESRec\xspace}
\newcommand{\itemAA}{outlier items\xspace}
\newcommand{\CapitemAA}{Outlier Items\xspace}
\newcommand{\fullmodel}{\textbf{C}onversation \textbf{E}nhanced \textbf{S}equential \textbf{Rec}ommendation\xspace}
\newcommand{\pseudofull}{semantic-based pseudo interaction construction\xspace}
\newcommand{\maskfull}{dual alignment outlier items masking\xspace}

\newcommand{\aka}{\emph{a.k.a.,}\xspace}

\newcommand{\ignore}[1]{}

\title{\model: Constructing Pseudo Interactions \\ for Sequential Recommendation via Conversational Feedback}
\author{
  \textbf{Yifan Wang\textsuperscript{1}},
  \textbf{Shen Gao\textsuperscript{1}},
  \textbf{Jiabao Fang\textsuperscript{2}},
  \textbf{Rui Yan\textsuperscript{3}},
\\
  \textbf{Billy Chiu\textsuperscript{4}},
  \textbf{Shuo Shang\textsuperscript{1}}
\\
\\
  \textsuperscript{1}School of Computer Science and Technology, University of Electronic Science and Technology of China\\
  \textsuperscript{2}School of Computer Science and Technology, Shandong University\\
  \textsuperscript{3}School of Artificial Intelligence, Wuhan University\\
  \textsuperscript{4}Department of Computing and Decision Sciences, Lingnan University\\
\\
}
\begin{document}
\maketitle
\begin{abstract}
Sequential Recommendation Systems (SRS) have become essential in many real-world applications. 
However, existing SRS methods often rely on collaborative filtering signals and fail to capture real-time user preferences, while Conversational Recommendation Systems (CRS) excel at eliciting immediate interests through natural language interactions but neglect historical behavior. 
To bridge this gap, we propose \model, a novel framework that integrates the long-term preference modeling of SRS with the real-time preference elicitation of CRS. 
We introduce \pseudofull, which dynamically updates users' historical interaction sequences by analyzing conversational feedback, generating a pseudo-interaction sequence that seamlessly combines long-term and real-time preferences. 
Additionally, we reduce the impact of outliers in historical items that deviate from users' core preferences by proposing \maskfull, which identifies and masks such items using semantic-collaborative aligned representations. 
Extensive experiments demonstrate that \model achieves state-of-the-art performance by boosting strong SRS models, validating its effectiveness in integrating conversational feedback into SRS\footnote{Code is available at https://github.com/NNNNyifan/CESRec}.
\end{abstract}

\section{Introduction}

Sequential Recommendation Systems (SRS) are pivotal in various applications, such as e-commerce~\cite{zhou2018deep} and streaming platforms~\cite{pan2023understanding}, by providing personalized item recommendations based on users' historical interaction sequences~\cite{fang2020deep}. 
Recently, large language models (LLMs) have demonstrated remarkable reasoning capabilities~\cite{mann2020language,zhang2022automatic}, making them promising method for enhancing recommendation tasks. 
Several studies~\cite{liao2024llara,bao2023tallrec} have demonstrated the superiority of directly applying LLMs to sequential recommendation tasks. 
In contrast, Conversational Recommendation Systems (CRS) employ natural language interactions to inquire about user preferences and predict personalized item recommendations~\cite{friedman2023leveraging,mysore2023large}. 
\begin{figure}[t]
\centering
  \includegraphics[width=\linewidth]{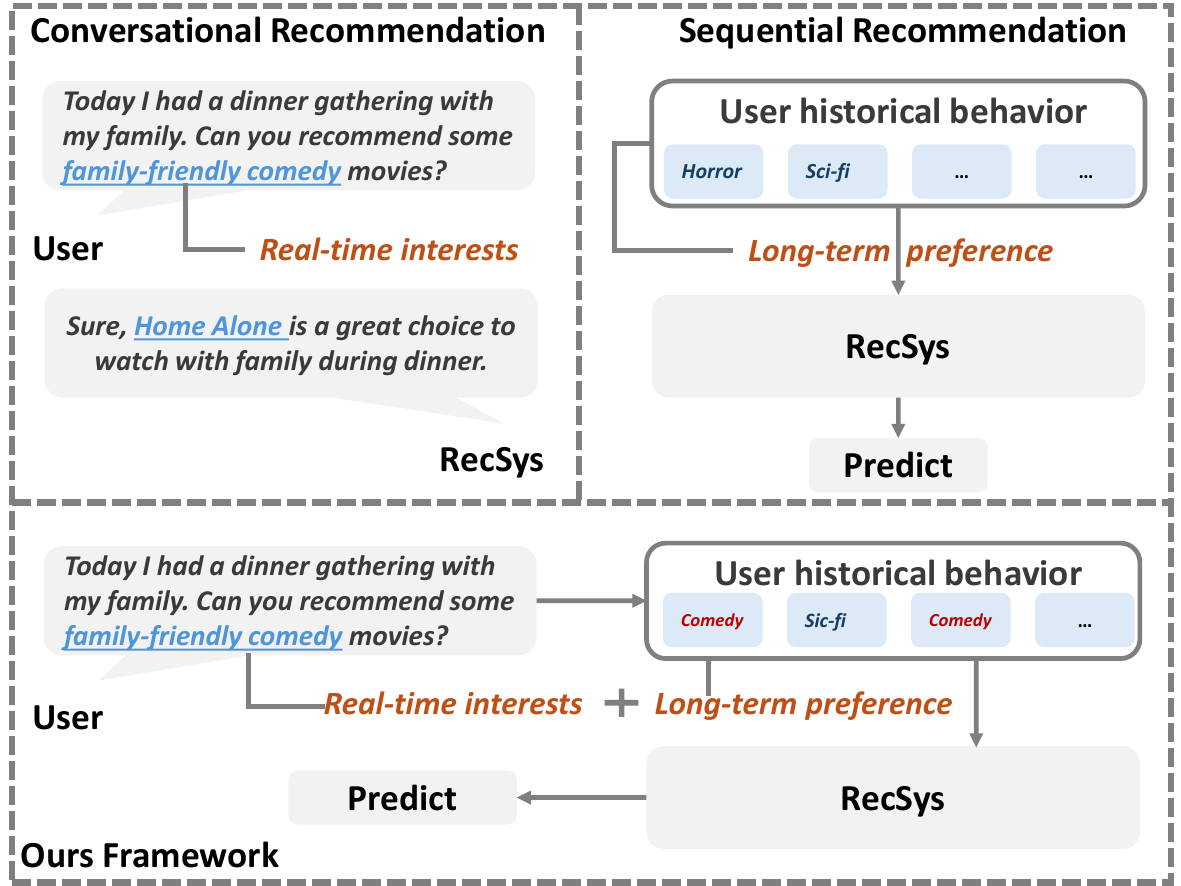}
  \caption{Comparison of sequential recommendation, conversational recommendation, and our \model, which leverage advantage of conversational recommendation to enhance sequential recommendation.}
  \label{fig:intro}
\end{figure}
However, existing SRS methods usually rely on collaborative filtering signals while neglecting the rich semantic information associated with items.
A significant limitation of these approaches is their inability to capture users' real-time interests, as immediate preferences are not dynamically reflected in the behavior sequence.
Conversely, while CRS methods excel at capturing immediate interests through natural language conversations, they typically fail to incorporate historical interaction sequences into their frameworks~\cite{zhou2020towards,lei2020interactive,he2023large,feng2023large}.
Consequently, the first challenge lies in dynamically integrating the \textit{long-term preference modeling} of SRS with the \textit{real-time interests modeling} facilitated by natural language feedback in CRS.

In this paper, we propose \fullmodel (\model).
To address the first challenge, we introduce \textbf{\pseudofull}, a novel method that directly refines the historical interaction sequence based on users' natural language feedback. 
Specifically, this approach analyzes users' natural language feedback to model their current preferences and refines their historical interaction sequence, generating a \textit{pseudo-interaction sequence} that seamlessly integrates both long-term and real-time preferences.
Next, we use the pseudo-interaction sequence as input to SRS, which effectively combines the collaborative filtering signals of SRS with the semantic signals derived from natural language feedback. 
This enables accurate recommendations based on natural language interactions without requiring extensive modifications to existing SRS-based systems, ensuring seamless integration and enhanced user experience.

Since historical interaction sequences often contain items that deviate substantially from users' main preferences, such as mistakenly clicked items or transient interests, as observed in many recent studies~\cite{lin2023self,wang2021denoising}, these outliers can adversely affect the modeling of user behavior.
The inclusion of these items can negatively influence the LLM's modeling of user behavior, potentially misleading the construction of the pseudo-interaction sequence. 
For example, if a user's primary preference is horror films, the inclusion of a comedy movie in the interaction sequence might lead the LLMs to utilize ``horror-comedy'' films to construct the pseudo-interaction sequence, rather than a pure horror film. 
In this work, we refer to such items as \textbf{\itemAA}.
Therefore, the second challenge is how to accurately identify these \itemAA and mask them in the interaction sequence to minimize their impact on the generation of the pseudo-interaction sequence.

To address this, we propose \maskfull, a method that accurately identifies \itemAA from the user's historical interaction sequence based on semantic-collaborative aligned representations and subsequently masks these items. 
Specifically, we leverage LLMs to obtain semantic embeddings of items and extract collaborative representations from the SRS model. 
We then introduce a dual alignment mechanism to derive hybrid item representations, which simultaneously capture co-occurrence relationships and semantic information among items. 
Based on these hybrid representations, we identify items that substantially deviate from the user's core preferences, ensuring precise masking while preserving the integrity of the user's historical behavior sequence. 
The experimental results demonstrate that our \model can boost the performance of several state-of-the-art SRS models in terms of HR and NDCG, 
which verifies that our \model can effectively integrates long-term preferences with real-time interests through natural language feedback.
\noindent The main contributions of this work are as follows:

\noindent $\bullet$ We propose \model, which combines the advantage of real-time natural language feedback with the efficiency of learning user preferences from historical behavior. 

\noindent $\bullet$ We introduce \pseudofull method to refine user historical interaction sequences by leveraging user natural language feedback. 

\noindent $\bullet$ We propose \maskfull method to optimize item selection during the sequence refinement process. 

\noindent $\bullet$ Extensive experiments demonstrate that our proposed \model achieves state-of-the-art performance by boosting the performance of several strong SRS models.

\section{Related Work}
\paragraph{Sequential Recommendation}
Sequential recommendation aims to predict the next item that aligns with a user's preferences based on their historical interaction sequence~\cite{fang2020deep,li2023strec, li2023automlp}. 
Traditional sequential recommendation models capture user preferences by leveraging item co-occurrence relationships. 
To model complex sequential patterns, CNN-based~\cite{tang2018personalized} and GNN-based~\cite{he2020lightgcn} methods have been introduced. 
Additionally, transformer-based approaches, such as SASRec~\cite{kang2018self} and BERT4Rec ~\cite{sun2019bert4rec}, have been developed to capture long-term dependencies between arbitrary items.
However, most of these methods primarily model user preferences based on long-term interaction histories, making it challenging to effectively capture dynamic shifts in user interests. 
As a result, they struggle to reflect users' real-time preferences within interaction sequences, leading to recommendations that may not accurately align with users' immediate interests.
\paragraph{LLMs for Recommendation}
Large Language Models (LLMs) have demonstrated remarkable capabilities across various domains. 
Recent research explores hybrid approaches to enhance sequential recommendation by integrating traditional Recsys with LLMs~\cite{dai2023uncovering,geng2022recommendation,hou2024large,bao2023tallrec}. 
\cite{liao2024llara} combines ID-based embeddings and textual features via hybrid prompting, while \cite{rajput2023recommender} introduces generative retrieval using semantic ID decoding. \cite{liu2024large} leverages LLMs to generate item embeddings, and \cite{hu2024enhancing} aligns ID embeddings with text via a projector module.
However, they do not fully exploit the rich semantic information contained in users' conversational feedback, limiting their ability to dynamically adapt recommendation strategies based on real-time user preferences.
\section{Problem Definition}

In this paper, we follow the problem definition commonly used in sequential recommendation tasks~\cite{hu2024enhancing}. 
Given a user $u \in \mathcal{U}$, where $\mathcal{U}$ represents the set of all users, and a historical interaction sequence $\mathcal{I}(u) = \{v^{(u)}_1, v^{(u)}_2, \dots, v^{(u)}_{N_u}\}$, the model aims to predict the next item the user is likely to interact with based on $\mathcal{I}(u)$. 
Here, $v^{(u)}_i$ denotes the $i$-th item interacted by user $u$, and all items belong to the item set $\mathcal{V}$. 
The sequence length of $\mathcal{I}(u)$ is denoted by $N_u$.
\section{\model}

\subsection{Overview}
\begin{figure*}[bht]
\centering
  \includegraphics[width=\linewidth]{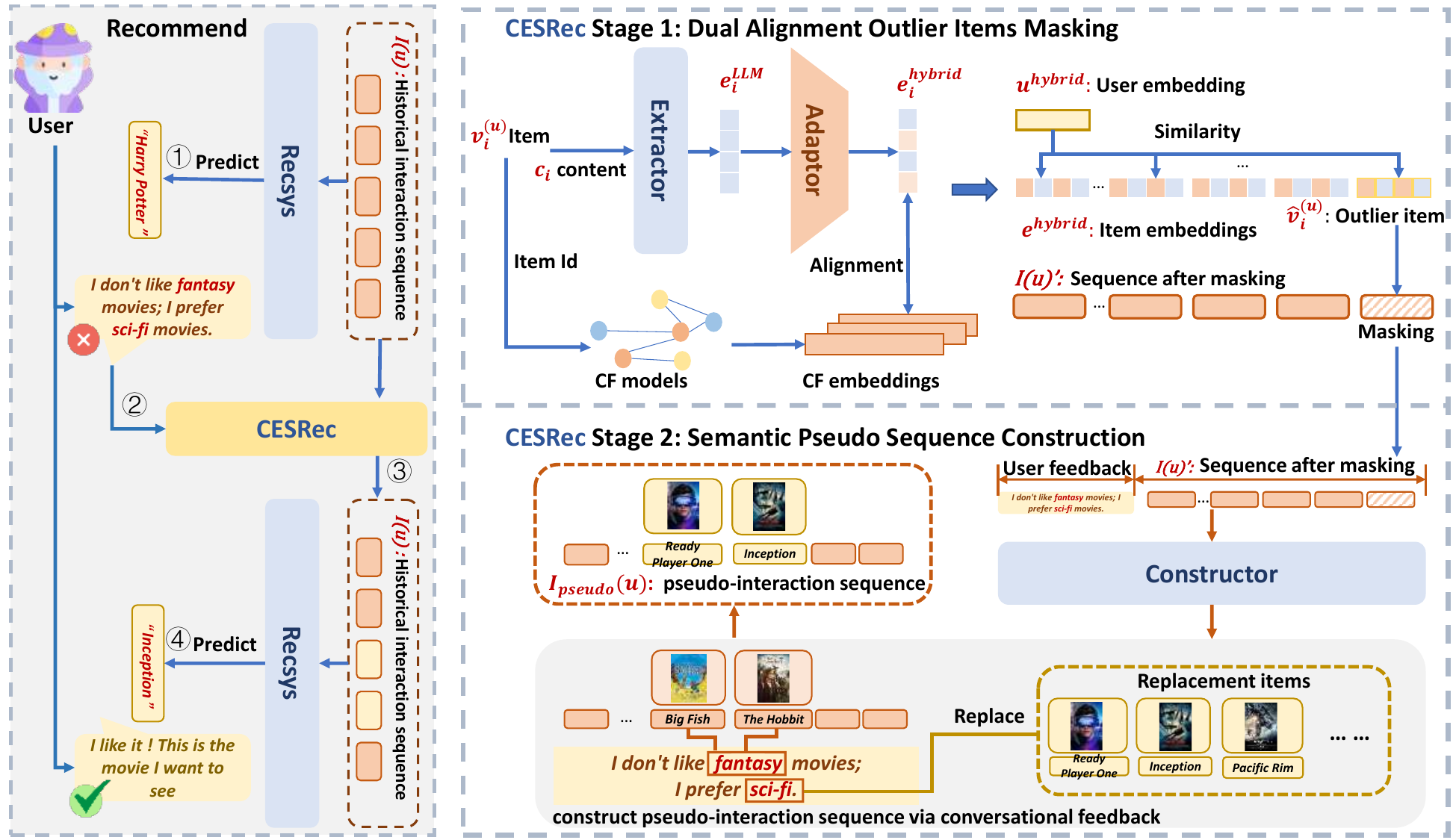}
  \caption{Overview of \model. In our proposed framework, we first employ the conventional sequential recommendation method (\aka Recsys) to predict an item based on the user's historical interaction sequence. Next, our \model refines the interaction sequence by constructing the pseudo-interaction sequence and masking the \itemAA. Finally, we employ Recsys to give a new recommendation by using the refined sequence.}
  \label{fig:model}
\end{figure*}

In this section, we show the details of the \fullmodel (\model), which is illustrated in Figure~\ref{fig:model}. 
The proposed model consists of two main components: Semantic Pseudo Sequence Construction and Dual Alignment \CapitemAA Masking.
The \textbf{Semantic Pseudo Sequence Construction} module is designed to construct the pseudo-interaction sequence by refining the historical interaction sequence via users' conversational feedback.
Subsequently, the \textbf{Dual Alignment \CapitemAA Masking} module further enhances the refinement process by identifying and masking items that deviate from the user's core preferences.

\subsection{Dual Alignment \CapitemAA Masking}
In the process of constructing a semantic-based pseudo-interaction sequence, the model leverages the users historical sequences to capture their core preferences and selects appropriate replacement items based on natural language feedback. 
However, during the modification of the original interaction sequence, items in the historical sequence that deviate from the user's core preferences can interfere with the LLM's modeling of user behavior. 
This misalignment can introduce bias, potentially leading to the inappropriate replacement of items. 
In this work, we refer to such items as \itemAA.
To address this issue, we propose a dual-alignment \itemAA masking method to ensure that such deviating items are appropriately masked.

According to a recent study~\cite{sheng2024language}, LLMs can implicitly encode user preference information, and items sharing similar content tend to exhibit similar semantic embeddings.
Based on this observation, we extract item embeddings from LLMs, which are rich in semantic information.
Given an item $v_i^{(u)}$ with content information $c_i$ such as title, we employ an LLM to obtain the semantic embeddings $e_i^{LLM}$:
\begin{equation}
e_i^{LLM} = \text{Extractor}(c_i),
\end{equation}
where $\text{Extractor}(\cdot)$ refers to the LLM tokenizer and model layers, and we utilize the output of the last hidden layer $e_i^{LLM}$ as the semantic embedding.

Relying solely on semantic embeddings to identify \itemAA may compromise the integrity of the user’s historical behavior sequence, thereby limiting the effectiveness of SRS in accurately modeling user preferences. 
We introduce a trainable adapter to align the semantic embeddings derived from LLMs with the collaborative signals typically used in SRS. 
This adapter is specifically trained to fuse the positional influence and co-occurrence information while utilizing semantic embeddings for masking:
\begin{equation}
{e_i^{\text{hybrid}}} = \text{Adapter}(\theta_{collab}; e_i^{LLM}),\label{equ:adapter}
\end{equation}
where $e_i^{\text{hybrid}} \in \mathbb{R}^{d_r}$ denote the hybrid embedding that combines both semantic and collaborative information through an adaptation module. 
The adaptation is performed by $\text{Adapter}_{\theta}: \mathbb{R}^{d_l} \rightarrow \mathbb{R}^{d_r}$ with trainable parameters $\theta_{\text{collab}}$, which projects embeddings from the LLM's latent space ($d_l$-dimensional) to the Recsys's embedding space ($d_r$-dimensional). 
The detailed architecture and training procedure of the adapter are described in Appendix~\ref{app:dual-alignment}

Finally, to identify \itemAA in interactions, we rank items based on the similarity between user representation and each item.
We first obtain all the hybrid embeddings of all the user interacted items in $ \mathcal{I}(u)$, and fuse all the item representation as the user embedding $u^{\text{hybrid}}$:
\begin{equation}
    u^{\text{hybrid}} = \text{Fuse}(\{e_1^{\text{hybrid}}, e_2^{\text{hybrid}}, \dots, e_{N_u}^{\text{hybrid}}\}),
\end{equation}
where the $\text{Fuse}(\cdot)$ denotes the mean-pooling operator.
Then, we calculate the similarity between each item representation $e_i^{\text{hybrid}}$ and user representation $u^{\text{hybrid}}$.
\begin{equation}
s_i = \text{Similarity}(e_i^{\text{hybrid}},u^{\text{hybrid}}),
\end{equation}
where $s_i \in [0, 1]$ denotes the similarity score, and we employ the cosine similarity as the $\text{Similarity}(\cdot)$ function to measure the semantic gap between $e_i^{\text{hybrid}}$ and $u^{\text{hybrid}}$.
To identify \itemAA in interaction sequence, we rank items based on their similarity scores $s_i$.
The top $k$ items with the lowest similarity scores are considered as the \itemAA and will be subsequently masked from the user interaction sequence.

The input and output format of the final dual alignment \itemAA masking is as follows:
\begin{equation}
I(u)^\prime= \text{Dual-Alignment}(I(u)),
\end{equation} 
where ${I(u)}^\prime = \{v^{(u)}_1, \dots, v^{(u)}_{N_u-k}, \hat{v}^{(u)}_1, \dots, \hat{v}^{(u)}_k\}$ represents interaction sequence after masking, $\hat{v}^{(u)}_i$ represents the top $k$ items with the lowest similarity scores. 
Using these hybrid representations, we identify and mask the \itemAA that deviate from the user’s core preferences while preserving the integrity of their historical behavior sequence.
This optimization enables the \model to better concentrate on core preferences when constructing a semantic-based pseudo interaction sequence.

\subsection{Semantic Pseudo Sequence Construction}

To address the challenge of dynamically integrating long-term preference modeling of SRS with real-time interest modeling driven by natural language interactions in CRS, we propose a semantic-based pseudo sequence construction approach. 
This method leverages natural language feedback from users to directly capture their current preferences, and generates semantic-based pseudo sequence by incorporating current preferences to historical interaction sequence.
Specifically, we introduce a $constructor$ that constructs semantic-based pseudo interaction sequences based on user-provided feedback. 
Following the previous works~\cite{fang2024multi}, we ask the user for preference about the target item attributes.
\begin{equation}
\text{feedback}=\text{User-Interaction}(v_{rec}^{(u)},Attr_{\text{target}})
\end{equation}
where $v_{rec}^{(u)}$ represents the recommended item generated by an SRS with input $I(u)$, $Attr_{\text{target}}$ refers to attributes of the target item, and $\text{feedback}$ denotes a natural language feedback derived from the user that describes the user preference of the item attributes.
For instance, if the SRS recommends <Avatar> to the user, but the user prefers films directed by Christopher Nolan, the user may respond with feedback such as: \textit{``I don't like film directed by James Cameron; I prefer Christopher Nolan.''}

Next, the $\text{Constructor}$ integrates user feedback to iterative refine the historical interaction sequence $I^\prime(u)$ and generate the pseudo-interaction sequence $I_{\text{pseudo}}(u)$:
\begin{equation}
I_{\text{pseudo}}(u)=\text{Constructor}(I^\prime(u), \text{feedback}),
\end{equation}
where $I_{\text{pseudo}}(u)$ represents the pseudo-interaction sequence generated by the $\text{Constructor}$.
~\paragraph{Training of Constructor}
To achieve a synergistic integration of user long-term interests and real-time preferences, we fine-tune LLM as the Constructor.
The training objective for the Constructor is formulated as a sequence prediction task:
\begin{equation}
\mathcal{L}_{\text{seq}} = -\sum_{t=1}^{|I_{pseudo}^*(u)|} \log P_\Psi\big(v_t^* | v_{<t}^*, I(u), \text{feedback}\big)
\end{equation}
$I_{pseudo}^*$ denotes the pseudo-interaction sequence, which optimally combines the user's long-term interests and real-time preference reflected in user feedback.

To construct the training data for the constructor module, we generate the pseudo-interaction sequence by replacing items that no longer align with the user's current preferences.
During the training data construction process, we randomly sample items from the interaction sequence as ``\CapitemAA''. The target item, which reflects the user's current preference, serves as the ground truth for model training. The feedback derived from the transition between the \CapitemAA and the target item is utilized as the input feature for the model.
The training instruction is as follows:
\begin{tcolorbox}[
    title=Instruction of Constructor,
    label=tbox:feedback,
    colback=white,
    colframe=black!50,
    fonttitle=\bfseries,
    sharp corners
]
\small \textbf{Instruction:} Based on the preferences mentioned in the user feedback and the information about \texttt{<items>} contained in the historical interaction sequence, replace the \texttt{<items>} the user dislikes with \texttt{<items>} user may currently prefer.\\
\small \textbf{Input:} historical interaction sequence: \texttt{<sequence>}; user feedback: \texttt{<feedback>}.\\
\small \textbf{Output:} pseudo-interaction sequence:<pseudo sequence>
\end{tcolorbox}

Finally, after refining the interaction sequence of the user by the $\text{Constructor}$, we use the semantic pseudo interaction sequence $I_{\text{pseudo}}(u)$ as the input to the SRS to regenerate recommended items.
\begin{equation}
v^{(u)}_{N_u+1}=\text{SRS}(I_{\text{pseudo}}(u)),
\end{equation}
where SRS represents sequential recommendation models, $v^{(u)}_{N_u+1}$ represents the regenerated recommended item based on the semantic pseudo interaction sequence.
Since our proposed \model is model-agnostic, it can be seamlessly integrated with existing sequential recommendation models.

\section{Experimental Setup}

\subsection{Dataset and Evaluation Metric}
\begin{table}[h]
\centering
\small 
\resizebox{\linewidth}{!}{%
\begin{tabular}{l|cccc}
\toprule
\textbf{Dataset} & \textbf{\#User} & \textbf{\#Item} & \textbf{\#Review} & \textbf{\#Density}\\
\midrule
Video Games & 55,223 & 17,408 & 496,315 & 0.051628\%\\
Toys & 208,180 & 78,772 & 1,826,430 & 0.011138\% \\
MovieLens & 6,040 & 3,883 & 1,000,209 & 4.264680\%\\
\bottomrule
\end{tabular}%
}
\caption{Statistics of three datasets.}
\label{table:statistics_of_datasets}
\end{table}

\begin{table*}[t]
\centering
\resizebox{1\linewidth}{!}{
\begin{tabular}{l|l|cccc|l|cccc}
\toprule
\textbf{Dataset} & \textbf{Traditional Model} & \textbf{HR@5} & \textbf{NDCG@5} & \textbf{HR@10} & \textbf{NDCG@10} & \textbf{LLM-based Model} & \textbf{HR@5} & \textbf{NDCG@5} & \textbf{HR@10} & \textbf{NDCG@10} \\
\midrule
\multirow{3}{*}{Video Games} 
& SASRec & 0.590 & 0.4629 & 0.717 & 0.5042 & LLaRA & 0.270 & 0.2277 & 0.360 & 0.2558 \\
& +\model-LLaMA2 & 0.633 & 0.4847 & 0.725 & 0.5144 & +\model-LLaMA2 & 0.380 & 0.3097 & \textbf{0.450} & 0.3316 \\
& +\model-LLaMA3 & \textbf{0.646} & \textbf{0.4923} & \textbf{0.745} & \textbf{0.5242} & +\model-LLaMA3 & \textbf{0.380} & \textbf{0.3254} & 0.440 & \textbf{0.3445} \\
\midrule
\multirow{3}{*}{Movielens} 
& SASRec & 0.757 & 0.5688 & 0.866 & 0.6045 & LLaRA & 0.170 & 0.1416 & 0.210 & 0.1542 \\
& +\model-LLaMA2 & \textbf{0.824} & \textbf{0.6076} & 0.882 & \textbf{0.6264} & +\model-LLaMA2 & 0.260 & 0.2192 & 0.310 & 0.2347 \\
& +\model-LLaMA3 & 0.810 & 0.5996 & \textbf{0.886} & 0.6244 & +\model-LLaMA3 & \textbf{0.280} & \textbf{0.2348} & \textbf{0.330} & \textbf{0.2508} \\
\midrule
\multirow{3}{*}{Toys} 
& SASRec & 0.431 & 0.3173 & 0.537 & 0.3509 & LLaRA & 0.420 & 0.3957 & 0.430 & 0.3986 \\
& +\model-LLaMA2 & 0.472 & 0.3376 & 0.557 & 0.3647 & +\model-LLaMA2 & 0.500 & 0.4671 & 0.590 & 0.4955 \\
& +\model-LLaMA3 & \textbf{0.478} & \textbf{0.3408} & \textbf{0.557} & \textbf{0.3659} & +\model-LLaMA3 & \textbf{0.500} & \textbf{0.4671} & \textbf{0.600} & \textbf{0.4993} \\
\bottomrule
\end{tabular}
}
\caption{Performance on three datasets. We apply our proposed \model on two strong SRS: SASRec and LLaRA, and we implement \model based on two LLM backbones: LLaMA2 and LLaMA3.}
\label{table:main_result}
\end{table*}

\begin{table}[h]
\centering
\resizebox{1\linewidth}{!}{
\begin{tabular}{l|l|c|c|c|c}
\toprule
\textbf{Dataset} & \textbf{Method} & \textbf{HR@5} & \textbf{NDCG@5} & \textbf{HR@10} & \textbf{NDCG@10} \\
\midrule
\multirow{4}{*}{\textbf{Video Games}} 
& \texttt{+\model-LLaMA3} & \textbf{0.646} & \textbf{0.4923} & \textbf{0.745} & \textbf{0.5242} \\
& \texttt{+\model w/o d.a.} & 0.634 & 0.4849 & 0.723 & 0.5136 \\
& \texttt{+\model w/o c.} & 0.610 & 0.4711 & 0.723 & 0.5077 \\
& \texttt{SASRec} & 0.590 & 0.4629 & 0.717 & 0.5042 \\
\midrule
\multirow{4}{*}{\textbf{Movielens}} 
& \texttt{+\model-LLaMA3} & \textbf{0.810} & \textbf{0.5996} & \textbf{0.886} & \textbf{0.6244} \\
& \texttt{+\model w/o d.a.} & 0.805 & 0.5940 & 0.880 & 0.6186 \\
& \texttt{+\model w/o c.} & 0.774 & 0.5766 & 0.866 & 0.6061 \\
& \texttt{SASRec} & 0.757 & 0.5688 & 0.866 & 0.6045 \\
\midrule
\multirow{4}{*}{\textbf{Toys}} 
& \texttt{+\model-LLaMA3} & \textbf{0.478} & \textbf{0.3408} & \textbf{0.557} & \textbf{0.3659} \\
& \texttt{+\model w/o d.a} & 0.468 & 0.3354 & 0.557 & 0.3638 \\
& \texttt{+\model w/o c.} & 0.443 & 0.3222 & 0.530 & 0.3501 \\
& \texttt{SASRec} & 0.431 & 0.3173 & 0.537 & 0.3509 \\
\bottomrule
\end{tabular}
}
\caption{Performance of ablation models. We conduct ablation study on SASRec+\model.}
\label{table:ablation_result}
\end{table}

We conduct experiments on two commonly used recommendation datasets, Video Games and Toys, constructed from the Amazon review datasets~\cite{ni2019justifying}.
We also employ the MovieLens datasets~\cite{harper2015movielens} which is a widely adopted dataset for sequential recommendation tasks, which contains user interactions with movies. 
Statistics are shown in Table~\ref{table:statistics_of_datasets}.

We adopt two widely used metrics to evaluate the performance: Normalized Discounted Cumulative Gain (NDCG@K) and Hit Rate (HR@K) with K=5,10.
We select 100 non-interacted items to construct the candidate set, ensuring the inclusion of the correct subsequent item. 

\subsection{Implementation Detail}
To evaluate our method, we employ gpt-4o-mini as a user simulator to provide natural language feedback (details in Appendix~\ref{app:user simulation}). 

We employ LoRA technique to fine-tune the LLM as the constructor module.

For the sequential recommendation method, SASRec~\cite{kang2018self}, we train the model on all three datasets using the Adam optimizer~\cite{kingma2014adam} for 200 epochs, with a learning rate of 0.001 and a batch size of 256.
For the LLM-based recommendation method, LLaRA~\cite{liao2024llara}, the original configuration selects the top-ranked item from the candidate set as the recommendation result. 

To ensure consistency with our experimental setup, we adopt the ranking method from ~\cite{wang2024llm4dsr}, which ranks the candidate items based on the cosine similarity between item embeddings and the output embeddings of LLaRA.
In our \model, we mask one item in three datasets.
We implement our \model using two LLMs as the backbone: LLaMA-2-7b~\cite{touvron2023LLaMA} and LLaMA-3-8b~\cite{dubey2024LLaMA}.
And we use the same user simulator as the previous conversational recommendation studies~\citet{fang2024multi} when training and evaluating the models.

\subsection{Baselines}

We conducted experiments using two strong SRS backbones: (1) \textbf{SASRec}~\cite{kang2018self} is a widely used sequential recommendation model that employs a self-attention mechanism to effectively capture relationships between items within a user's interaction sequence. 
(2) \textbf{LLaRA}~\cite{liao2024llara} is an LLM-based recommendation model that utilizes a hybrid prompting approach, combining ID-based and text-based representations of items as input. This model aims to enhance recommendation accuracy by integrating both structured and unstructured data sources.

\section{Experimental Results}

\subsection{Main Results}
We evaluate the performance of our proposed \model and baseline methods on three datasets using four evaluation metrics.
As shown in Table~\ref{table:main_result}, SASRec+\model and LLaRA+\model consistently outperform their corresponding base SRS model (\aka SASRec and LLaRA) across all datasets and metrics.
This demonstrates that the semantic-based pseudo interaction sequences, which incorporate users' current feedback, enable recommendation models to more effectively capture users' real-time preferences.

Secondly, \model demonstrates improved performance when leveraging larger LLMs as the backbone, suggesting that more powerful LLMs possess the stronger capability to accurately model user preferences and select relevant replacement items.

\subsection{Ablation Study}
To validate the effectiveness of each module, we compare the performance of the following variants of \model-LLaMA3 on the SASRec backbone:
(1) \textbf{\model w/o d.a.}: we solely employ user conversational feedback to construct pseudo interaction sequences and remove the \textbf{d}ual \textbf{a}lignment from \model.
(2) \textbf{\model w/o c.}: we only leverage dual alignment method to mask \itemAA and do not construct pseudo sequence.
The results, as shown in Table~\ref{table:ablation_result}, demonstrate that all modules in the model contribute to enhancing sequential recommendation. 
The superior performance of \model-LLaMA3 over \model w/o d.a. indicates that the dual alignment \itemAA masking method enables \model to concentrate on user's main preference, and construct semantic pseudo sequences that better align with user preferences.

\subsection{The Influence of LLM Backbones}
\begin{table}[t]
\centering
\resizebox{1\linewidth}{!}{
\begin{tabular}{l|cccc}
\toprule
\textbf{LLM Backbones} & \textbf{HR@5} & \textbf{NDCG@5} & \textbf{HR@10} & \textbf{NDCG@10} \\
\midrule
SASRec & 0.590 & 0.4629 & 0.717 & 0.5042 \\
+\model-Qwen2.5-3B-Instruct & \textbf{0.649} & \textbf{0.4929} & 0.717 & 0.5146 \\
+\model-LLaMA2-7B-Instruct & 0.633 & 0.4847 & 0.725 & 0.5144 \\
+\model-Mistral-7B-Instruct-v0.3 & 0.646 & 0.4906 & 0.734 & 0.5188 \\
+\model-Qwen2.5-7B-Instruct & 0.644 & 0.4911 & 0.732 & 0.5195 \\
+\model-LLaMA3-8B-Instruct & \underline{0.646} & \underline{0.4923} & \textbf{0.745} & \textbf{0.5242} \\
\bottomrule
\end{tabular}
}
\caption{Performance comparison of various LLM backbones on the Video Game dataset.}
\label{tab:results}
\end{table}
To assess the influence of varying LLM backbones on recommendation accuracy, we evaluated diverse LLM architectures and scales on Video Games dataset.
Employing the proposed \model framework, the small language model (SLM) Qwen2.5-3B-Instruct shows improved recommendation performance in resource-constrained settings, highlighting \model's effectiveness under limited resources.
Its fewer hidden layers compared to LLMs reduce semantic loss when mapping from the LLM’s latent space to the Recsys’s embedding space, enabling more accurate masking of \itemAA.
For models of comparable size, Qwen2.5-7B-Instruct and Mistral-7B-Instruct-v0.3 consistently surpass Llama-2-7B-Instruct, suggesting their generated pseudo-interaction sequences more accurately reflect real-time user interests.

\subsection{The Impact of Historical Interaction Sequence Length}
\begin{figure}[h]
\centering
  \includegraphics[width=1\linewidth]{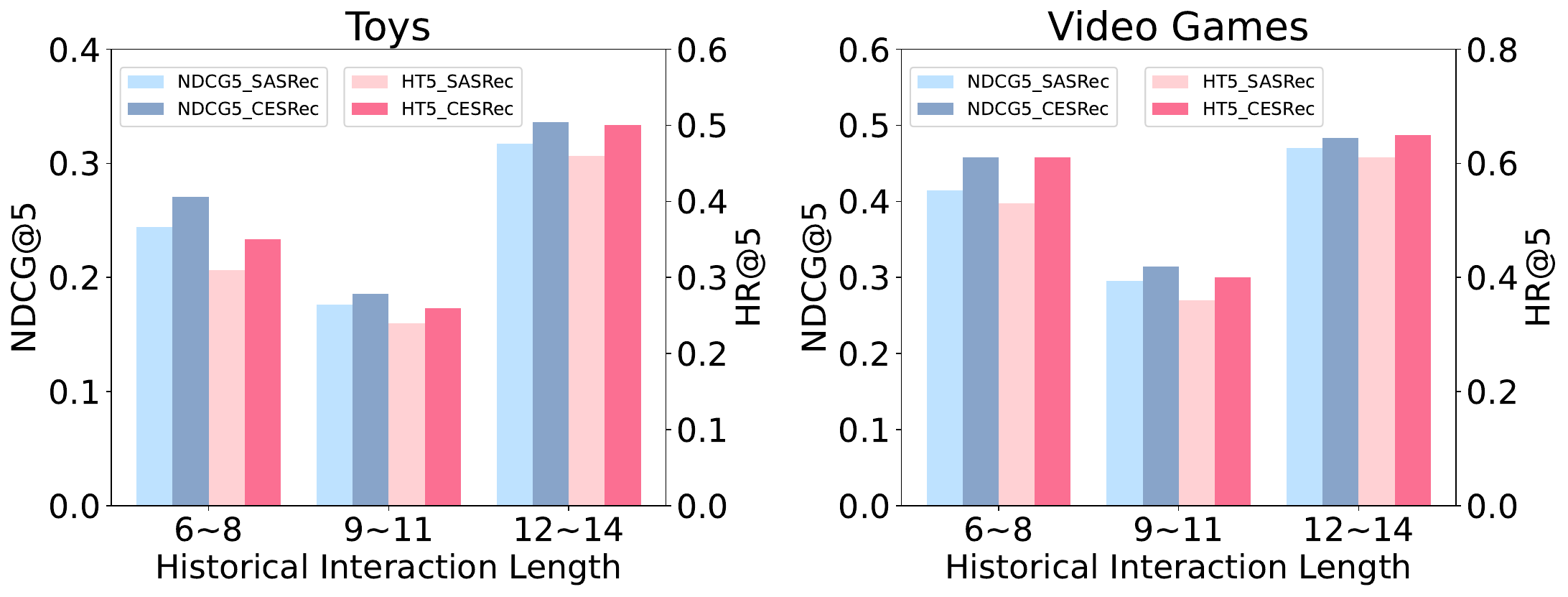}
  \caption{Performance of using different lengths of the historical interaction sequence.}
  \label{fig:seqlen_compare}
\end{figure}

To investigate the impact of historical interaction sequence length, we evaluate model performance using different sequence lengths in terms of HR@5 and NDCG@5 on the Toys and Video Games datasets. 
As shown in Figure~\ref{fig:seqlen_compare}, the results demonstrate that our proposed \model consistently outperforms the baseline \texttt{SASRec} across all three sequence length ranges. 
This demonstrates the robustness of our model in effectively handling historical interaction sequences of varying lengths, further confirming its adaptability in diverse recommendation scenarios.
\subsection{Analysis of Interaction Numbers}

\begin{figure}[h]
\centering
  \includegraphics[width=\linewidth]{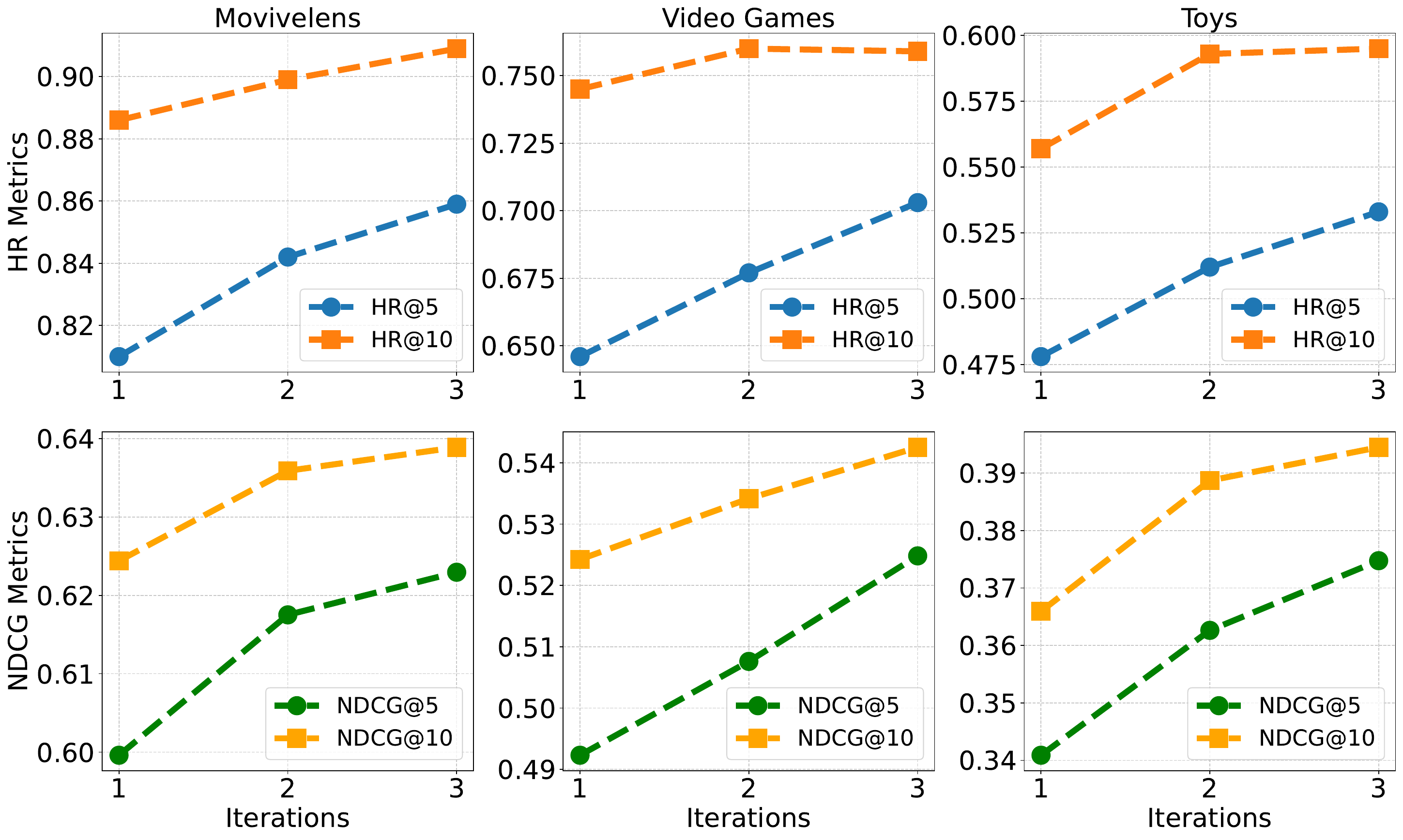}
  \caption{The influence of different numbers of user natural language feedback.}
  \label{fig:iterative_num}
\end{figure}

We further investigate the impact of the number of natural language feedback of \model-LLaMA3, based on SASRec. 
As illustrated in Figure~\ref{fig:iterative_num}, as the number of interactions between users and the \model-LLaMA3 increases, the performance of the Recsys consistently improves. 
The HR@K and NDCG@K metrics (with K=5, 10) demonstrate a steady upward trend across all three real-world datasets. 
This indicates that as users provide more feedback, the Recsys becomes increasingly effective at capturing users' real-time interests. 
By constructing semantic-based pseudo interaction sequences that reflect these interests, the system generates recommendations that better align with users' current preferences. 

The improvement in both HR and NDCG metrics demonstrate the model's enhanced ability to both identify and effectively rank relevant items, yielding more user-centric recommendations.

\subsection{Analysis of Masking \CapitemAA}
\begin{figure}[h]
\centering
  \includegraphics[width=1\linewidth]{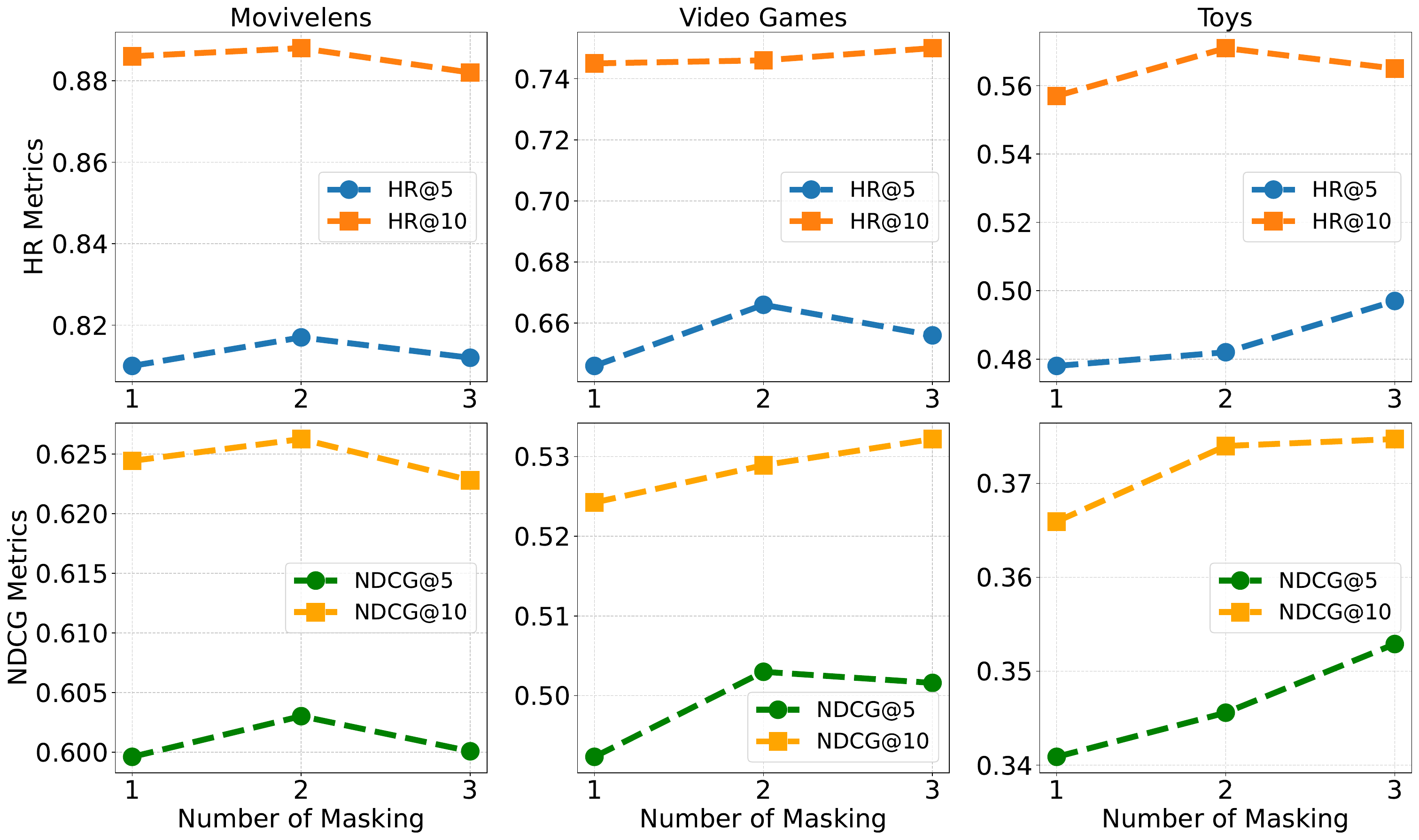}
  \caption{The impact of masking different numbers of \itemAA.}
  \label{fig:denoise_analysis}
\end{figure}
We further investigated the impact of the number of masked \itemAA on the performance of \model. 
The results show that for the MovieLens and Video Games datasets, the model achieves optimal performance when the number of masked items is set to 2. 
Beyond this threshold, performance begins to decline as the number of masked items increases. 
This decline can be attributed to the fact that excessive masking reduces the length of the user's historical sequence, leading to a loss of valuable information regarding user preferences. 
Consequently, the model struggles to accurately capture user behavior and predict items that align with these preferences.
In contrast, for the Toys dataset, the model's performance improves as the number of masked items increases.  
This trend can be attributed to the higher sparsity of the Toys dataset compared to other two datasets, as shown in Table~\ref{table:statistics_of_datasets}. 
With greater sparsity, the items in the constructed sequences exhibit more variability, and as the model adjusts these sequences based on user feedback expressed in natural language, the impact on the recommendation outcomes becomes more notable. 
Therefore, by masking items that deviate from the user's preferences, the model can concentrate on the most relevant interactions, resulting in improved performance.

\subsection{Case Study}

\begin{figure}
\centering
  \includegraphics[width=0.95\linewidth]{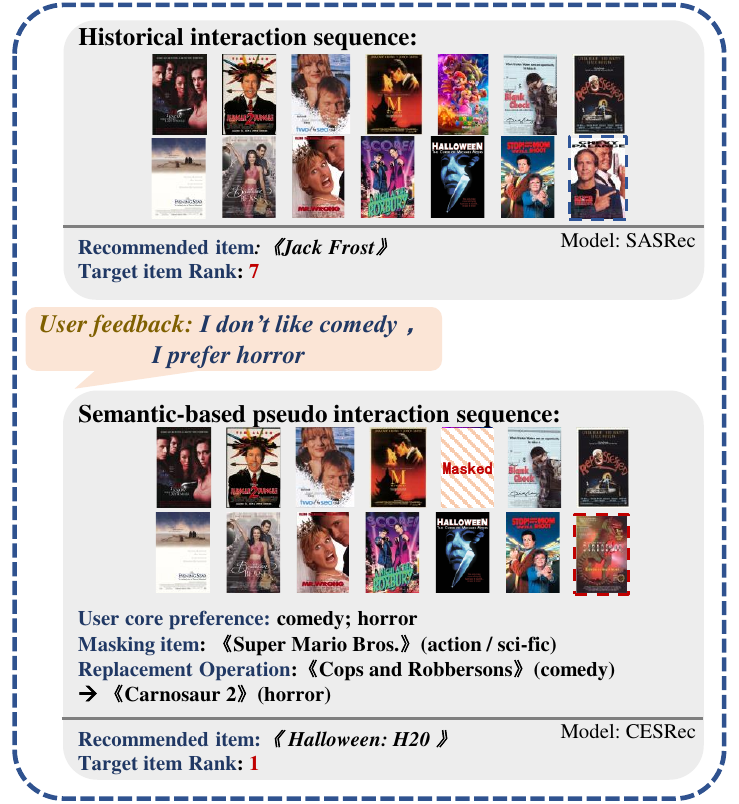}
  \caption{A case study of \model.}
  \label{fig:good_case}
\end{figure}

To intuitively validate the effectiveness of our proposed \model, we randomly select an example from MovieLens dataset, as shown in Figure~\ref{fig:good_case}.
The detail user's historical interactions with movies are shown in Appendix~\ref{app:case study}. 
Given this sequence as input, SASRec generates ``Jack Frost'' as a recommended item by capturing the co-occurrence relationships between movies. 
However, ``Jack Frost'' is a comedy film, which does not align with the user's current preference for horror films.
To encourage the model's focus on the user's core interests, we employ the dual alignment \itemAA masking method. 
This method masks the ``Super Mario Bros.'', which belongs to the action/animation genre and deviates from user's core preference for horror films. 
Thus, the model can better align with user's primary interests and improve recommendation accuracy.
This masking process enables the \model to better concentrate on the user's core preferences.
Since ``Jack Frost'' is inconsistent with the user's preference, \model constructs a semantic-based pseudo-interaction sequence incorporating the user's conversational feedback: ``I don't like comedy; I prefer horror.''.
During this process, \model replaces ``Cops and Robbersons (comedy)'' with ``Carnosaur 2 (horror)'' to reinforce the user's stated preference.
Ultimately, based on this refined interaction sequence, \model predicts ``Halloween: H20'' as the recommended item.

\section{Conclusion}

In this paper, we proposed \fullmodel (\model), a novel framework that seamlessly integrates the long-term preference modeling of SRS with the real-time preference elicitation of CRS. 
By leveraging users' natural language feedback, \model dynamically refines historical interaction sequences to generate pseudo-interaction sequences that capture both long-term preferences and real-time interests. 
Additionally, the \maskfull method addresses the challenge of \itemAA in historical sequences by accurately identifying and masking items that deviate from users' core preferences. 
Extensive experiments on three real-world datasets demonstrate that \model enhances the performance of SOTA SRS models, achieving superior results in terms of HR and NDCG metrics.

\clearpage
\section*{Limitations}
Our method relies on user conversational feedback to dynamically refine the historical interaction sequence, aiming to better align with the user's real-time preferences.
However, if the user's feedback is expressed in a vague, ambiguous, or unclear manner, the model may fail to capture the user's real-time preferences accurately, leading to the generation of an imprecise pseudo-interaction sequence, which in turn affects the recommendation performance.
In future work, we will investigate more sophisticated dialogue mechanisms that can effectively guide users to articulate their latent preferences.

\section*{Ethical Considerations}
The research conducted in this paper centers on investigating the effectiveness of leveraging LLMs to bridge the gap between conversational recommendation and sequential recommendation. 
Our work systematically benchmarks LLMs under various real-world scenarios and evaluates their performance. 
In the process of conducting this research, we have adhered to ethical standards to ensure the integrity and validity of our work.
To minimize potential bias and ensure fairness, we employ the same prompts and experimental setups as those used in existing publicly accessible and freely available studies.
We have made every effort to ensure that our research does not harm individuals or groups and does not involve any form of deception or misuse of information.

% \section*{Acknowledgments}

\bibliography{custom}
\section{Appendix}
\subsection{Theoretical Foundation for Dual Alignment Strategies} \label{app:dual-alignment}
Our dual alignment strategy maps LLM-derived semantic embeddings $\mathbf{e}_i^{\text{LLM}}$ to the recommendation space via an MLP-based \textbf{Adapter}, minimizing the MSE loss between hybrid embeddings $\mathbf{e}_i^{\text{hybrid}}$ and collaborative embeddings $\mathbf{e}_i^{\text{collab}}$.

\begin{equation}
\begin{split}
\mathbf{e}_i^{\text{hybrid}} &= \text{Adapter} (\mathbf{e}_i^{\text{LLM}}) \\
&= W_2 \cdot \text{GELU}(W_1 \cdot \mathbf{e}_i^{\text{LLM}} + b_1) + b_2
\end{split}
\end{equation}
\begin{equation}
L_{\text{align}} = \| \mathbf{e}_i^{\text{hybrid}} - \mathbf{e}_i^{\text{collab}} \|_2^2
\end{equation}
This ensures $\mathbf{e}_i^{\text{hybrid}}$ fuses semantic and collaborative signals.

\subsection{User Simulation}\label{app:user simulation}
Inspired by the principle that \textit{"while users may not always articulate their preferences, they can reliably identify what they dislike,"} our model initially generates recommendations based on the user’s historical interactions. 
If the user expresses dissatisfaction with a recommended item, they provide natural language feedback highlighting the specific features they find undesirable.

Following the work~\citet{fang2024multi}, we adopt a similar user simulation mechanism.
To mitigate potential data leakage, the user simulator is not directly exposed to the target item itself; instead, it receives only descriptive information about the target item.
The user simulator is prompted to provide feedback as follows: \textit{"You are a user interacting with a recommender system. Based on the information about your <target item> and the <recommended item> provided by the recommender, give feedback to the recommender."}

\subsection{Influence of Feedback Types} \label{app:feedback types}

The polarity of user feedback (positive or negative) influences the interaction sequence reconstruction process. The table~\ref{tbox:feedback} illustrates \model's adaptive capability in responding to varying feedback polarities.

\subsection{Incorporate with Language-based Baseline}
We integrate the \model into the Recformer~\cite{10.1145/3580305.3599519} and evaluated this approach on the Video Game dataset.
% (Li et al. 2023 [3]) 
The experimental results are shown in Table~\ref{tab:results_recformer}, demonstrating that after inserting our \model, Recformer shows improvements in NDCG@10, Recall@10, NDCG@50, Recall@50, and MRR metrics. Additionally, we selected different LLMs as baselines. Due to differences in semantic understanding capabilities, the enhancement effects vary across LLM baselines. In the experiments enhancing Recformer, the best performance was achieved using Qwen2.5-7b as the baseline.

\begin{table}[t]
\centering
\resizebox{1\linewidth}{!}{%
\begin{tabular}{lccccc}
\toprule
Model & NDCG@10 & Recall@10 & NDCG@50 & Recall@50 & MRR \\
\midrule
Recformer & 0.0323 & 0.0625 & 0.04489 & 0.1125 & 0.02906 \\
+ \model + Mistral v0.3-7B & 0.0323 & 0.0625 & 0.05089 & 0.1375 & 0.03053 \\
+ \model + LLaMA3-8B & 0.0368 & 0.0750 & 0.05859 & 0.1625 & 0.03321 \\
+ \model + Qwen2.5-7B & \textbf{0.0411} & \textbf{0.0875} & \textbf{0.06561} & \textbf{0.1875} & \textbf{0.03518} \\
\bottomrule
\end{tabular}%
}
\caption{The Recformer experimental results on the Video Game dataset.}
\label{tab:results_recformer}
\end{table}

\subsection{Alternative Similarity Functions}
We adopted cosine similarity for semantic alignment, as prior work~\cite{sheng2024language} demonstrated its effectiveness in evaluating semantic proximity.
To assess robustness, we replaced cosine similarity with Euclidean distance on the Video Game dataset. As shown in table~\ref{tab:sim_performance}, the results confirmed that cosine similarity outperformed alternatives in identifying outlier items and improving the precision of the recommendations.
\begin{table}[h]
\centering
\resizebox{1\linewidth}{!}{
    \begin{tabular}{lcccc}
        \toprule
        Model & HT@5 & NDCG@5 & HT@10 & NDCG@10 \\
        \midrule
        SASRec & 0.590 & 0.4629 & 0.717 & 0.5042 \\
        +\model-llama3-L2 & 0.644 & 0.4941 & 0.722 & 0.5191 \\
        +\model-llama3-cosine & 0.646 & 0.4923 & 0.745 & 0.5242 \\
        \bottomrule
    \end{tabular}
    }
    \caption{Performance of Different Similarity Function}
    \label{tab:sim_performance}
\end{table}

\subsection{Inference Latency}
To evaluate practical deployment feasibility, we measure the average latency of LLaMA3-8B-\model and Qwen2.5-7B-\model when performing two key operations: (1) outlier item masking and (2) pseudo-sequence construction, across three benchmark datasets:

\begin{table}[ht]
\centering
\resizebox{\linewidth}{!}{%
\begin{tabular}{lccccc}
\toprule
Model & Component & Video Game & Toys & Movielens \\
\midrule
\multirow{3}{*}{\model-LLaMA3-8B} 
& Outlier Items Masking & 0.1407s & 0.1597s & 0.1757s \\
& Pseudo Sequence Construction & 10.4124s & 7.6398s & 9.1576s \\
& Total Time & 10.5532s & 7.7996s & 9.3334s \\
\midrule
\multirow{3}{*}{\model-Qwen2.5-7B}
& Outlier Items Masking & 0.1520s & 0.1806s & 0.1722s \\
& Pseudo Sequence Construction & 6.8974s & 7.9245s & 4.0569s \\
& Total Time & 7.0494s & 8.1051s & 4.2292s \\
\bottomrule
\end{tabular}%
}
\caption{Inference Latency Comparison (Average Time)}
\label{tab:latency}
\end{table}

We calculated the average inference time across 10 samples. The real-time inference latency for outlier item masking ranges between 0.14-0.17s, while pseudo sequence construction constitutes the major portion of the inference delay.
Based on real-time user feedback, the total inference time for LLaMA3-8B to construct pseudo sequences and generate new recommendations ranges between 7-10s, while Qwen2.5-7B ranges between 4-8s.
Since different models have varying inference speeds, faster models can be selected based on strictly demanding production requirements. 

\subsection{Detail Information of Case study}\label{app:case study}
\begin{tcolorbox}[
    title=User Interaction Sequence of Case Study,
    label=tbox:feedback,
    colback=white,
    colframe=black!50,
    fonttitle=\bfseries,
    sharp corners
]
\begin{itemize}[leftmargin=*, noitemsep, topsep=0pt]  % Compact itemized list
    \item I Still Know What You Did Last Summer  
    \item Jungle 2 Jungle  
    \item Two if by Sea  
    \item M. Butterfly  
    \item Super Mario Bros  
    \item Blank Check  
    \item Repossessed  
    \item The Evening Star  
    \item The Beautician and the Beast  
    \item Mr. Wrong  
    \item A Night at the Roxbury  
    \item Halloween: The Curse of Michael Myers  
    \item Stop! Or My Mom Will Shoot  
    \item Cops and Robbersons  
\end{itemize}
\end{tcolorbox}

% \clearpage
\begin{tcolorbox}[
    title=Examples of bias migration analysis of \model,
    label=tbox:feedback,
    colback=white,
    colframe=black!50,
    fonttitle=\bfseries,
    sharp corners
]
\textbf{User Interaction Sequence:}
\medskip
\noindent

['Dumb and Dumber', 'The Hangover', 'Bridesmaids', 'Anchorman', 'The Exorcist', 'Hereditary', 'The Conjuring', 'A Nightmare on Elm Street', 'The Babadook', 'It', 'Superbad', 'Step Brothers']
\medskip
\noindent\makebox[\linewidth]{\dotfill}
\medskip
\textbf{Positive Feedback ("I like comedies")}
\medskip

Replaces <The Conjuring (horror)> to <Dogma (comedy)> to amplify comedy recommendations.

\medskip
\textbf{Sequence reconstructed by \model:}
\medskip

["Dumb and Dumber", "The Hangover", "Bridesmaids", "Anchorman", "The Exorcist", "Hereditary", "Dogma", "A Nightmare on Elm Street", "The Babadook", "It", "Superbad", "Step Brothers"]
\medskip
\noindent\makebox[\linewidth]{\dotfill}
\medskip
\textbf{Negative Feedback ("I dislike comedies")}
\medskip

Replaces <Bridesmaids (comedy)> to <Sleepy Hollow (horror)> to align with implicit horror preferences.

\medskip
\textbf{Sequence reconstructed by \model:}
\medskip

["Dumb and Dumber", "The Hangover", "Sleepy Hollow", "Anchorman", "The Exorcist", "Hereditary", "The Conjuring", "A Nightmare on Elm Street", "The Babadook", "It", "Superbad", "Step Brothers"]
\end{tcolorbox}

% \appendix

% \section{Example Appendix}
% \label{sec:appendix}

% This is an appendix.

\end{document}